\documentclass[10pt,conference]{IEEEtran}
\IEEEoverridecommandlockouts

\usepackage{colortbl}
\usepackage{graphicx}
\usepackage{amsmath}
\usepackage{amssymb}
\usepackage{booktabs}
\usepackage{graphicx}
\usepackage{epstopdf}
\usepackage{flushend}
\usepackage{multirow}
\usepackage[caption=false]{subfig}
\usepackage{amsmath}
\usepackage{amssymb}
\usepackage{booktabs}
\usepackage{listings}
\usepackage{svg}
\usepackage{xcolor}
\usepackage{threeparttable}
\usepackage{enumerate}
\usepackage{float}
\usepackage{pgfplots}
\tikzset{global scale/.style={
    scale=#1,
    every node/.append style={scale=#1}
  }
}
\usetikzlibrary{spy}
\usepgfplotslibrary{groupplots}
\pgfplotsset{width=0.6\linewidth,compat=1.9}

\usepackage{hyperref}
\hypersetup{
colorlinks=true,
linkcolor=black,
}
\usepackage{epstopdf}
\usepackage{gensymb}
\usepackage{textcomp}
\usepackage[ruled,linesnumbered]{algorithm2e}

\usepackage{amsmath}
\usepackage{bm}
\def\degree{${}^{\circ}$}
\usepackage[bottom=1in,top=0.75in,left=0.62in,right=0.62in]{geometry}

\DeclareRobustCommand*{\IEEEauthorrefmark}[1]{%
    \raisebox{0pt}[0pt][0pt]{\textsuperscript{\footnotesize\ensuremath{#1}}}}
\newcommand{\etal}{{et al}. }
\newcommand{\ie}{{i}.{e}. }
\newcommand{\eg}{{e}.{g}. }

\begin{document}
\title{\fontsize{16pt}{0pt} \textbf{A Cooperative Perception System Robust to Localization Errors}}

\author{{Zhiying Song\IEEEauthorrefmark{1}, Fuxi Wen\IEEEauthorrefmark{*}\IEEEauthorrefmark{1}, 
Hailiang Zhang\IEEEauthorrefmark{1} and Jun Li\IEEEauthorrefmark{1}}
\thanks{
\IEEEauthorrefmark{1}
School of Vehicle and Mobility, Tsinghua University, Beijing 100084, China. Email: \{song-zy21, zhanghl22\}@mails.tsinghua.edu.cn, \{wenfuxi, lijun1958\}@tsinghua.edu.cn. \IEEEauthorrefmark{*} Corresponding author.

}
}

\maketitle
\begin{abstract}
Cooperative perception is challenging for safety-critical autonomous driving applications.
The errors in the shared position and pose cause an inaccurate relative transform estimation and disrupt the robust mapping of the Ego vehicle. We propose a distributed object-level cooperative perception system called OptiMatch, in which the detected 3D bounding boxes and local state information are shared between the connected vehicles. To correct the noisy relative transform, the local measurements of both connected vehicles (bounding boxes) are utilized, and an optimal transport theory-based algorithm is developed to filter out those objects jointly detected by the vehicles along with their correspondence, constructing an associated co-visible set. A correction transform is estimated from the matched object pairs and further applied to the noisy relative transform, followed by global fusion and dynamic mapping. Experiment results show that robust performance is achieved for different levels of location and heading errors, and the proposed framework outperforms the state-of-the-art benchmark fusion schemes, including early, late, and intermediate fusion, on average precision by a large margin when location and/or heading errors occur.
\end{abstract}


\section{Introduction}
\label{section:intro}

Automated driving relies on the accurate perception of the surrounding vehicles and dynamic environment. However, automated vehicles are limited by the physical capabilities  (\eg, resolution and detection range) of the onboard sensors, therefore connected and automated vehicles (CAV) becomes a promising paradigm in recent years. 

\begin{figure}[htbp]
    \centering
    \subfloat[]{
    \includegraphics[height=2.2in]{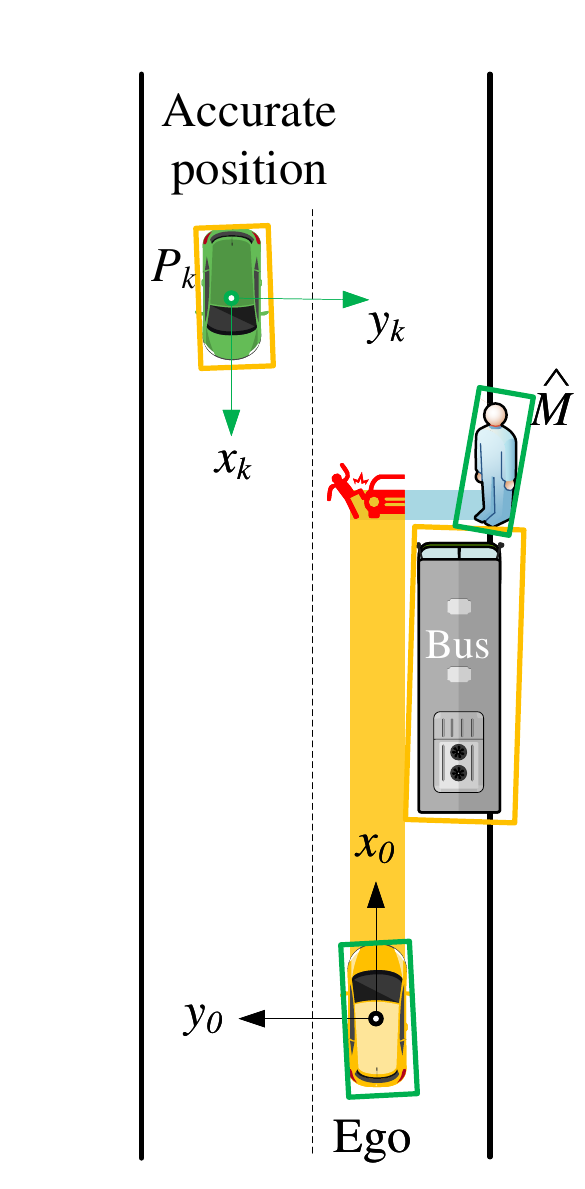}}
    \subfloat[]{
    \includegraphics[height=2.2in]{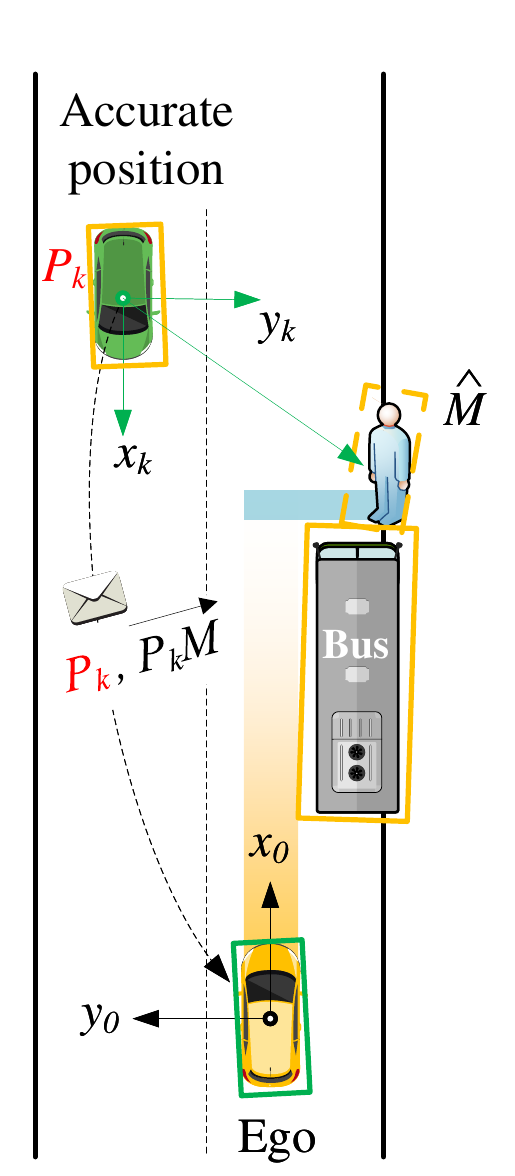}
    }
    \subfloat[]{
    \includegraphics[height=2.2in]{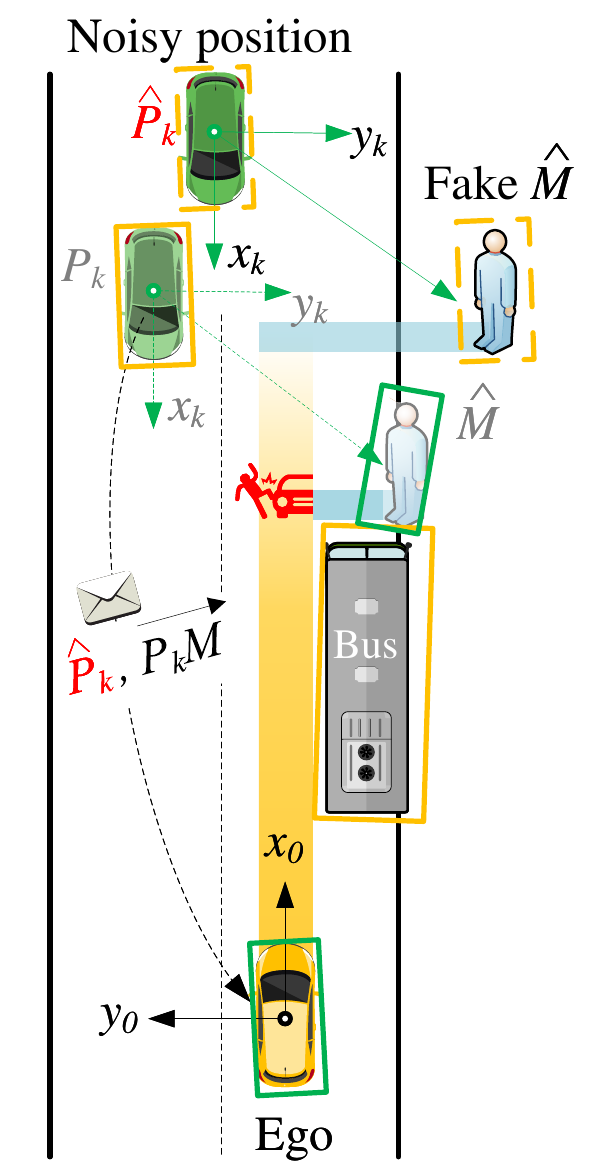}
    }
    \caption{\textbf{Illustration of cooperative perception.}  Colored rectangles represent perception of cooperative vehicles with corresponding colors. Solid lines indicate direct perception, and dashed lines show fused results. (a) {No cooperation.} The Ego  might crash the pedestrian $\hat{M}$ because of the occlusion of the bus.  (b) {Accurate cooperation.} CAV$_k$ sends its own accurate location $P_k$ and the relative location of $\hat{M}$ to the Ego. The crash might not happen.  (c) {Inaccurate cooperation.} CAV$_k$ sends a noisy location $\hat{P_k}$ to Ego, a fake $\hat{M}$ will appear from Ego's perspective. The crash might still happen.}
    \label{fig:coordinates}
\end{figure}
CAVs are connected via vehicle-to-vehicle (V2V) or vehicle-to-everything (V2X) communications and sense the surrounding environments through multi-agent cooperation. The effect of cooperative driving is illustrated in Fig.\ref{fig:coordinates}.
In practice, the effectiveness of cooperative perception depends on two aspects:  1) real-time and reliable data transmission within the limited network bandwidth, and 2) robust information fusion under highly dynamic and noisy environments.

The primary bottleneck for cooperative perception is the sharing of precise data with low latency and low communication burden  \cite{thandavarayan2019analysis}. 
Generally, sharing raw data provides the best performance because the least amount of information is lost. But it can easily overload the communication network with a large amount of real-time data transmission.
As a trade-off, features extracted from the
raw data by deep neural networks can reduce the amount of data to be shared and simultaneously maintain a good data fusion performance.
To further reduce the communication load, sharing fully processed
data, such as the information of the detected objects, requires fewer communication resources.
In this paper, we fuse data from different CAVs at the object level,  sharing the 3D bounding boxes, location, and pose information between the CAVs. 
This minimizes the burden on the communication network and allowing for rapid processing. Most importantly,  it is independent of onboard sensors and general among multiple scenarios.
The second challenge for cooperative perception is robust information fusion in highly dynamic and noisy environments. 
For cooperative perception at the object level, 
data received from other CAVs must first be converted to the Ego frame. 
In reality, the transforms are estimated from sensor measurements with limited resolution and accuracy, such as global positioning system  (GPS), real-time kinematic  (RTK), and inertial measurement unit  (IMU).
In most cases, the estimated transform shared among the CAVs are inaccurate, disrupting the robust mapping of the Ego vehicle in the process of cooperation.

This paper focuses on the above challenges, and the main contributions are summarized as follows:

\begin{itemize} 

\item A distributed V2V-based cooperative perception system is proposed, optimal transport theory is introduced to automatically correct inaccurate vehicle location and heading measurements using only object-level bounding boxes.
\item Experiments show that the proposed system outperform the state-of-the-art framework on two benchmark datasets in terms of robustness when location or heading errors occur, demonstrating the potential of simple object-level fusion to handle dynamic errors.
\item The proposed system gives a general solution independent of the type and model of onboard sensors, which can be easily extended to the vehicle-to-everything-based scenarios, and the proposed system transmits only object-level information, providing a low-cost solution with a low communication burden and easy implementation.
\end{itemize} 

The rest of the paper is organized as follows: In Section \ref{sec:related work}, the related work on cooperative perception and optimal transport is introduced. The problem is formulated in Section \ref{sec:problem}. Section \ref{sec:approach} contains the proposed object-level cooperative perception framework and detailed algorithms. Experimental results and discussion are presented in Section \ref{section:experiments}.

\section{Related work}
\label{sec:related work}
\subsection{Cooperative Perception}
Recent studies mainly focus on the aggregation of multi-agent information to improve the average precision of perception results. Arnold \etal evaluated the performance of early, and late fusion, as well as their hybrid combination schemes in driving scenarios using infrastructure sensors \cite{2020Cooperative}.
F-Cooper introduced feature-level data fusion that extracts and aggregates the feature map of the raw sensor data by deep learning networks and then detects objects on the fused feature map  \cite{Chen2019}.
V2VNet aggregated the feature information received from nearby vehicles and took the downstream motion forecasting performance into consideration  \cite{Wang2020}.
OPV2V released the first large-scale simulated V2V cooperation dataset and presented a benchmark with $16$ implemented models, within which we implement our models \cite{Xu2022a}. However, these existing studies are vulnerable to location and pose errors that are common and inevitable in real-world applications. 

FPV-RCNN tried to introduce a location error correction module based on key-point matching before feature fusion to make the model more robust  \cite{Yuan2022}. 
Vadivelu \etal  proposed a deep learning-based framework to estimate potential errors  \cite{vadivelu2020learning}, but they rely on feature-level fusion, which requires high computational capacity and is not general among different scenarios. 
Gao \etal proposed a graph matching-based method to identify the correspondence between the cooperative vehicles and can be used to promote the robustness against spatial errors \cite{gao2021regularized}. They formulated the problem as a non-convex constrained optimization problem and developed a sampling-based algorithm to solve it, however, the problem is difficult to solve and time-consuming, which hinders its application in the real-world.
In this paper, we try to take these errors into account and design an efficient and robust object-level cooperative perception framework.

\subsection{Optimal Transport Theory}
The optimal transport (OT) theory has been widely used in the assignment problem in various fields.
In the field of intelligent vehicles, Hungarian algorithm is one of the most popular variations of optimal transport methods and has been widely used to match two targets for its effectiveness and low complexity $O(n^3)$. For instance, Cai \etal used it to
assign vehicles to the generated goals in a formation to get
least lane changing overall \cite{cai2019multi}. 
For the perception problem, Sinkhorn's matrix scaling algorithm  \cite{sinkhorn1967diagonal} is more powerful for its high efficiency on the graphic processing unit (GPU) since Cuturi smoothed the classical optimal transport problem with an entropic regularization term in 2013  \cite{cuturi2013sinkhorn}. This makes the GPU available for the OT problem and accelerates its calculation much more than conventional methods. 
In recent years, OT with Sinkhorn has shown strong performance on several vision tasks with the rapid development of GPU. For example,
Sarlin \etal  \cite{sarlin2020superglue} formulated the assignment of graph features as a differentiable OT problem and acheived state-of-the-art performance on image matching.
Qin \etal  \cite{qin2022geometric} applied OT theory on the point cloud registration problem and developed a method with 100 times acceleration with respect to traditional methods. 
For the efficiency of OT and the Sinkhorn algorithm, it is deployed to find the object correspondences between the observation of the Ego and CAVs.

\section{Problem Formulation}
\label{sec:problem}
We consider a distributed cooperative perception scenario, where any cooperative CAV can share the local state and the information of the detected objects with the Ego vehicle. Let $\mathcal{X}=\{\bm{o}_i,i=1,2,..,m\}$ be the object set detected by the Ego vehicle and $\mathcal{Y}=\{\bm{o}_j, j=1,2,..,n\}$ be the object set detected by the CAV. For object $i$, it is represented as a 6D vector $\bm{o}_i = \begin{bmatrix}
      \bm{x}_i^T,
       \bm{\theta}_i^T
     \end{bmatrix}^T$, 
where $\bm{x}_i\in\mathbb{R}^3$ and $\bm{\theta}_i\in\mathbb{R}^3$ are the 
 3D position and orientation, respectively.

Cooperative fusion is to transform $\mathcal{Y}$ into the Ego frame and aggregate it with $\mathcal{X}$. However, errors presented in the state of both connected vehicles cause an inaccurate relative transform estimation, which is to be corrected in this paper.
The first challenge is to determine the co-visible region and associate co-visible objects, given the local state of Ego vehicle and CAV, as well as noisy measurements $\mathcal{X}$ and $\mathcal{Y}$, provided that the co-visible objects set $\mathcal{M}$ is achievable.
The second problem is to estimate a transform $\mathcal{F} $ defined as the function of rotation matrix $\mathbf{R}\in SO (3)$ and translation vector $\mathbf{t}\in \mathbb{R}^{3}$ between objects in the $\mathcal{X}$ and $\mathcal{Y}$ to approach the accurate spatial transform.  
It can be formulated as the following optimization problem
\begin{equation}
\begin{aligned}
    \min_{\mathcal{F}} \ \ \ & \sum_{  (i,j)\in {\mathcal{M}}}  \left|\left|
    \bm{x}_i
    -\mathcal{F}(\bm{y}_j)
    \right|\right|^2 \\
\end{aligned}
\label{eq:transform}
\end{equation}
where ${\bm{x}}_i$ denotes the position vector of $ {\bm{o}}_i \in \mathcal{X}$ (similar as ${\bm{y}}_j$ to ${\bm{o}}_j\in\mathcal{Y}$), and 
$ (i,j)$ is a possible object pair representing the same target. Operator $\mathcal{F}(*)$ is defined as $
\mathcal{F}(*) =\mathbf{R}\cdot(*)+\mathbf{t}
$.
The third task is to complete the fusion using the estimated transform to maximize the perception capacity of the Ego.

\section{Proposed Method}
\label{sec:approach}
The proposed fusion framework consists of four submodules: {preprocess}, {co-visible object association}, {optimal transform estimation}, {global fusion and dynamic mapping}. 

\subsection{Preprocess}
Multi-agent cooperation relies on the transform between agents. In the proposed system, the Ego calculates the relative transform $\mathcal{F}^{(1)}=\mathbf{R}^{(1)}\cdot(*)+\mathbf{t}^{(1)}$ using the position and pose data of both Ego and cooperative CAV. Its rotation and translation component is
\begin{equation}
\mathbf{R}^{(1)} = \mathbf{R}_x(\Delta \theta)\cdot
\mathbf{R}_y(\Delta \psi) \cdot
\mathbf{R}_z(\Delta \phi)
\end{equation}
and
\begin{equation}
    \mathbf{t}^{(1)}=\mathbf{p}_{\text{cav}}-\mathbf{p}_{\text{ego}}
\end{equation}
where $\Delta \theta, \Delta \psi, \Delta \phi$ are the relative Euler angle measurements of the Ego and cooperative CAV, and $\mathbf{R}_{a}(\theta)$ denotes the rotation of $\theta$rad with respect to the axis $a \in \{x,y,z \}$. $\mathbf{p}_{\text{cav}}$ and $\mathbf{p}_{\text{ego}}$ are the position vector of the two vehicles in the global frame.

$\mathcal{F}^{(1)}$ is used to transform the received objects set $\mathcal{Y}$ into Ego frame to obtain a new set of $\mathcal{Y}^{(1)}$,
\begin{equation}
    \mathcal{Y}^{(1)}= {\mathcal{F}^{(1)}}(\mathcal{Y})
\end{equation}
which is well-aligned with $\mathcal{X}$ if no error contained in the relative pisition and pose.
However, errors do exist, inevitably. 

The driving mode of the Ego vehicle is then determined (single vehicle or cooperative mode). The cooperative mode is boosted only if
a common field-of-view (FOV) for the Ego and the cooperative CAV exists, \ie, some objects are co-visible by both cooperative vehicles ($\mathcal{X}\cap \mathcal{Y}^{(1)} \neq \Phi$). 

\subsection{Co-visible Object Association}
 The critical step to solving (\ref{eq:transform}) is to estimate the common target set $\hat{\mathcal{M}}$ and assign pair-wise correspondence. It can be formulated as an optimal transport (OT) problem to minimize the transportation cost between the source   (points in $\mathcal{X}$) and target   (points in $\mathcal{Y}^{(1)}$).
For noisy measurements $\hat{\bm{o}}_i \in \mathcal{X}$, we want to assign at most a unique correspondence from  $\hat{\bm{o}}_j^{(1)} \in \mathcal{Y}^{(1)}$. Similar to the graph matching task  \cite{sarlin2020superglue}, the following two constraints should be satisfied: 1) a target in $\mathcal{X}$ can have at most one single correspondence in $\mathcal{Y}^{(1)}$; and 2) some targets will be unmatched because of different visions, occlusion, or detection errors. 

We define a cost matrix $\mathbf{C} \in \mathbb{R}^{m \times n}$,  to describe the transportation cost for association, with 
\begin{equation}
    \mathbf{C}_{i,j}=||\hat{\bm{x}}_i-\hat{\bm{y}}_j^{(1)}||_2.
\end{equation}
In order to propose a generalized formulation to handle the non-matched points, augmenting cost matrix $\overline{\mathbf{C}} \in \mathbb{R}^{(m+1) \times (n+1)}$ is constructed by appending a new row and column called dustbin, filled with a single hyper-parameter $\alpha \in \mathbb{R}$ \cite {DeTone_2018_CVPR_Workshops}, 
\begin{equation}
\overline{\mathbf{C}}(m+1,:) = \alpha \bm{1}_{n+1}^T 
   \text{ and } \overline{\mathbf{C} }(:,n+1) = \alpha \bm{1}_{m+1} 
\end{equation}
Therefore, for $\hat{\boldsymbol{o}}_i \in \mathcal{X}$, it is either matched to $\hat{\boldsymbol{o}}^{(1)}_j \in \mathcal{Y}^{(1)}$ or to the dustbin.

Once the cost matrix is defined, the task is to find the optimal assignment matrix  $\overline{\mathbf{P}} \in \mathbb{R}^{ (m+1) \times   (n+1)}$, where $\overline{\mathbf{P}}_{i,j}$ denotes the assignment probability on $\overline{\mathbf{C}}_{i,j}$, then we have the following modified form of the optimal transport problem:
\begin{equation}
\begin{aligned}
 \min_{\overline{\mathbf{P}}}\quad & \sum_{i,j} -\overline{\mathbf{P}}_{i,j} \overline{\mathbf{C}}_{i,j}\\
 {s.t.}\quad & \overline{\mathbf{P}} \bm{1}_{n+1} =\begin{bmatrix}
  \bm{1}_m^T,\;n
 \end{bmatrix}
^T, \\
& \overline{\mathbf{P}}^T \bm{1}_{m+1} =\begin{bmatrix}
  \bm{1}_n^T,\;m
 \end{bmatrix}
^T
\end{aligned}
\end{equation}
The equality constraint is a relaxation of the original element-wise inequality constraint ($\leq$), which allows each point to be matched with at most one point or dustbin, whereas each dustbin could be matched with all points at most. This relaxation makes the problem computationally efficient to solve using the Sinkhorn algorithm on GPU \cite{cuturi2013sinkhorn}.

Assignment matrix $\mathbf{P}$ is constructed by dropping the last row and column of $\overline{\mathbf{P}}$, points $\mathcal{X}_i$ and $\mathcal{Y}_j$ are associated as an object pair being added into $\hat{\mathcal{M}
}$, if 
$$
\mathbf{P}_{i, j} = \mathrm{argmax} \; \mathbf{P}  (i,:) = \mathrm{argmax } \; \mathbf{P}  (:,j)
$$
where $i \in [1,m]$ and $j \in [1,n]$.

\subsection{Optimal Transform Estimation}

The estimated association set $\hat{\mathcal{M}}$ might be inaccurate due to outliers or noisy bounding box information  (perception errors).
Random sampling techniques are utilized to enhance the robustness of the proposed method. We sample a few pairs one time rather than taking all into consideration to estimate the transform. The procedure can be repeated $n_s$ times or processed in parallel to find the best transform that maximizes the correct matching ratio. 

For the $s$th step, we randomly select a subset $\hat{\mathcal{M}}_{s}\subseteq \hat{\mathcal{M}}$, where $s \in \{1, 2, \cdots, n_s\}$.
Let $\mathbf{X}=[\hat{\bm{x}}_1,...,\hat{\bm{x}}_w]$ and $\mathbf{Y}^{(1)}=[\hat{\bm{y}}^{(1)}_1,...,\hat{\bm{y}}^{(1)}_w]$ be the associated measurements within $\hat{\mathcal{M}}_s$. 
$\bm{\mu}_x =\left(\sum_{i=1}^w\hat{\bm{x}}_i\right)/w$ and $\bm{\mu}_{y^{(1)}}=\left(\sum_{j=1}^w\hat{\bm{y}}_j^{(1)}\right)/w$ are further defined as the center of the measurements. 
Then the optimal transform in $\hat{\mathcal{M}}_s$  is given by \cite{Besl1992,dorst2005first}, 
\begin{equation}
    \mathcal{F}_s^{(2)}(*)=\mathbf{R}^{(2)}\cdot (*) + \bm{\mu}_x-\mathbf{R}^{(2)}\cdot \bm{\mu}_{y^{(1)}}
\end{equation}
where
$$
\mathbf{R}^{(2)}=\mathbf{U}\text{diag} \Big(1,1,\text{det}\left[\mathbf{U}\mathbf{V}^T\right]\Big)\mathbf{V}^T,
$$  
matrices $\mathbf{U}$ and $\mathbf{V}$ are obtainable by taking the Singular Value Decomposition on $
\mathbf{Y}^{(1)}\mathbf{X}^T=\mathbf{U}\Lambda \mathbf{V}^T$.

For each correspondence $ (\hat{\bm{x}}_i,\hat{\bm{y}}^{(1)}_i) \in  \hat{\mathcal{M}}$, it is regarded as an aligned pair if $||\hat{\bm{x}}_i-\mathcal{F}_s^{(2)}(\hat{\bm{y}}_i^{(1)})|| \leq  \tau$. 
Here threshold $\tau=0.25$m is set empirically because we find that a vanilla late fusion system without transform correction can handle the location error whose Gaussian standard deviation $\sigma_p \leq  0.2$m, and tighter $\tau$ decreases the effect of random sampling. 
For convenience, the set of the aligned point pairs is defined as $\hat{\mathcal{M}}_a^{(s)}$. Then correct matching ratio $\eta^{(s)}$ can be calculated by
\begin{equation}
\eta^{(s)} = {\text{card} \left( \hat{\mathcal{M}}_a^{(s)} \right)}/{\text{card} ( \hat{\mathcal{M}} )},
\end{equation}
where operator $\text{card}(*)$ denotes the number of elements in the set.
Optimal $s^{*}$ is determined by
\begin{equation}
s^* =\mathrm{argmax}_{s = 1, 2, \cdots, n_s}  \eta^{(s)}. 
\end{equation}

Finally, $\mathcal{F}^{(2)} = \mathcal{F}_{s^*}^{(2)}$ is obtained that can be further applied to global fusion, the solution of (\ref{eq:transform}) is 
\begin{equation}
    \hat{\mathcal{F}}=\mathcal{F}^{(2)}\circ \mathcal{F}^{(1)}
\end{equation}

\subsection{Global Objects Fusion and Dynamic Mapping}
All the objects lie under the Ego frame after imposing the correction transform $\mathcal{F}^{(2)}$ and noisy transform ${\mathcal{F}}^{(1)}$ on the objects detected by the cooperative CAV. 
\begin{equation}
\mathcal{Y}^{(2)}=\mathcal{F}^{(2)}(\mathcal{Y}^{(1)})
\end{equation}

It is typically done by Kalman filter or implemented under the Bayesian framework to merge those spatial-similar targets and remove those redundant objects in other vehicular tasks. 
These methods rely on initialization and require a few frames to warm up. 
Since our model estimates the relative transform frame by frame with high accuracy, Non-maximum suppression  (NMS) \cite{Xu2022a,Bodla2017} is more suitable to fuse the transformed objects. 
In general, NMS is integrated as the last step of the object detection algorithm requiring detection result (bounding boxes and corresponding classification confidence)
as input and exporting a list of objects meeting our requirements. As a result, the fused objects set is obtained by
\begin{equation}
    \mathcal{O}=\text{NMS}\left( \mathcal{X},\mathcal{Y}^{(2)} \right)
\end{equation}
For pseudo code of NMS, please refer to Figure 2 in \cite{Bodla2017}. 

\section{Experiments}
\label{section:experiments}

The proposed algorithm
and benchmarks are evaluated on the open-source dataset OPV2V \cite{Xu2022a}, which contains 73 scenes for V2V-based collaborative perception collected from the CARLA simulator \cite{dosovitskiy2017carla}.
The dataset is divided into four subsets: {Train}, {Validate}, {OPV2V-Test} and  {Digital Culver City} with $6764$, $1981$, $2170$ and $549$ frames, respectively. The last two subsets are used for evaluation, where data in {OPV2V-Test} has a similar distribution to the {Train} and {Validate}. {Digital Culver City} is specially built to narrow down the gap between the simulated and real-world traffic, which can be used to test the adaptability and portability of the proposed algorithm.
For fairness, all the compared methods use only LiDAR data and use the same backbone PointPillars  \cite{Lang_2019_CVPR} for object detection, all the models are implemented in PyTorch \cite{paszke2017automatic}  and run on NVIDIA GTX $3090$ GPU. 

\begin{figure}[htbp]
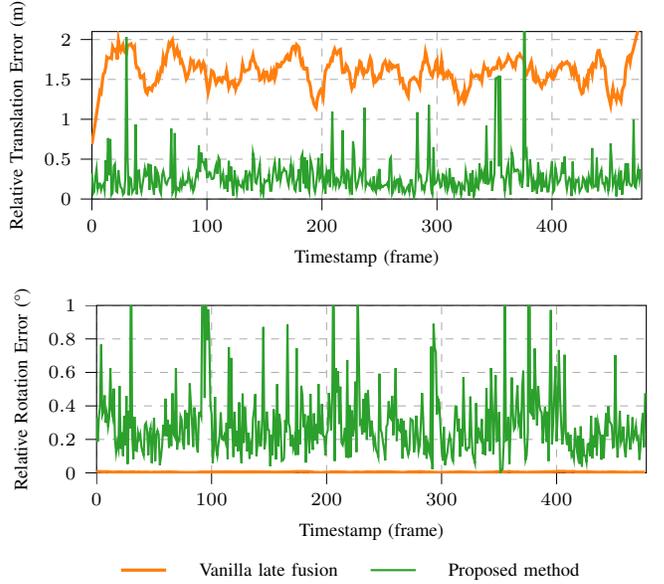

    \centering
    \subfloat{
    \input{materials/tikztex/RTE_compare.tex}
    }
    \vspace{-3mm}
\subfloat{
    \input{materials/tikztex/RRE_compare.tex}}
     \vspace{-2mm}
    \subfloat{
    \definecolor{mycolor1}{rgb}{0.00000,0.44700,0.74100}%
\definecolor{mycolor2}{rgb}{0.85000,0.32500,0.09800}%
\definecolor{mycolor3}{rgb}{0.92900,0.69400,0.12500}%
\definecolor{mycolor4}{rgb}{0.49400,0.18400,0.55600}%
\definecolor{mycolor5}{rgb}{0.46600,0.67400,0.18800}%
\definecolor{mycolor6}{rgb}{0.30100,0.74500,0.93300}%
\definecolor{mycolor7}{rgb}{0.63500,0.07800,0.18400}%
\definecolor{lightgray204}{RGB}{204,204,204}
\definecolor{darkgray176}{RGB}{176,176,176}
\definecolor{darkorange25512714}{RGB}{255,127,14}
\definecolor{forestgreen4416044}{RGB}{44,160,44}
\definecolor{darkblue}{RGB}{31,119,180}
\definecolor{darkred}{RGB}{210,0,0}

\begin{tikzpicture}[font=\scriptsize]
\node[draw=white,outer sep=0.5pt,inner sep=0pt] at (0,0) 
{\scriptsize
     \begin{tabular}{llll}
    \ref{woregistration} & Vanilla late
fusion &
    \ref{raw with regi}& Proposed method\\
    \end{tabular}
    };
\end{tikzpicture}
    }
    \caption{Relative error of estimated transform tested on 
  dataset {Digital Culver City} with position error ($\sigma_p=1$m)}
    \label{fig:rreandrte}
\end{figure}

The proposed system is compared with the mainstream early and late fusion methods, as well as the 
state-of-the-art  (SOTA) intermediate fusion strategies, in terms of the robustness against the position and heading errors of the CAVs.
As shown in  \cite{Xu2022}, intermediate fusion methods perform similarly on localization robustness. Here F-Cooper  \cite{Chen2019} and OPV2V \cite{Xu2022a} are selected as the representative of SOTA intermediate fusion models.  

To simulate the noisy measurements, white Gaussian noise is directly added to the ground truth position and pose of the cooperative CAV provided by the dataset.
Since only relative information between the cooperative CAV and the Ego matters in the cooperation process, such an operation is equivalent to adding noise both on the ground truth position and pose of the cooperative CAV and the Ego.
We denote the position and pose of the cooperative CAV at timestamp $t$ as 
$
  \bm{o}_t = [x,y,z,\theta,\psi,\phi]^T+\omega_t 
$, 
where $[x,y,z]$ define the ground truth location, and $[\theta,\psi,\phi]$ are ground truth pitch, roll, and heading  (yaw) angles, respectively. $\omega_t$ is zero-mean Gaussian noise with covariance matrix 
\begin{equation}
    \mathbf{Q}=\mathbf{E}(\omega_t\omega_t^T)=\text{diag}\left(\sigma_p^2,\sigma_p^2,\sigma_p^2,0,0,\sigma_\phi^2\right)
\end{equation}
 Here errors in pitch and roll are neglected for their slight impact on vehicle perception, and the standard deviations of noise in $[x,y,z]$ are chosen to be the same for simplicity, such an assumption does not influence the evaluation of robustness of the models.

 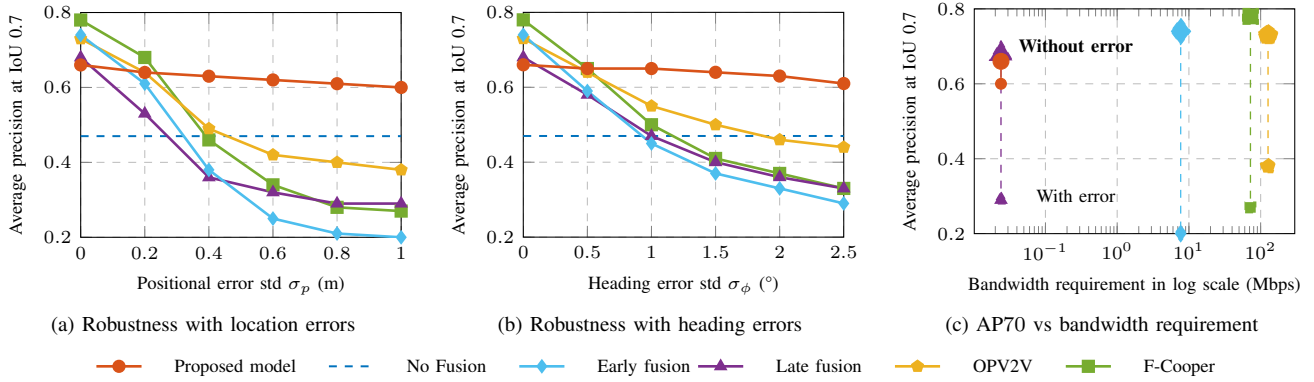
\begin{figure*}[htbp]
\centering
\subfloat[Robustness with location errors]{
\definecolor{mycolor1}{rgb}{0.00000,0.44700,0.74100}%
\definecolor{mycolor2}{rgb}{0.85000,0.32500,0.09800}%
\definecolor{mycolor3}{rgb}{0.92900,0.69400,0.12500}%
\definecolor{mycolor4}{rgb}{0.49400,0.18400,0.55600}%
\definecolor{mycolor5}{rgb}{0.46600,0.67400,0.18800}%
\definecolor{mycolor6}{rgb}{0.30100,0.74500,0.93300}%
\definecolor{mycolor7}{rgb}{0.63500,0.07800,0.18400}%
\definecolor{lightgray204}{RGB}{204,204,204}

\begin{tikzpicture}[font=\scriptsize]
\begin{axis}[
width=2.3in,
height=1.8in,
legend pos=outer north east,
legend style={fill opacity=0.8, draw opacity=1, text opacity=1, draw=lightgray204},
ymin=0.2,ymax=0.8,
xmin=0,xmax=1,
xlabel=Positional error std $\sigma_p$ (m),
ylabel=Average precision at IoU 0.7,
ymajorgrids=true,
xmajorgrids=true,
grid style=dashed,
]
\addplot[dashed,color=mycolor1,line width =0.75pt]
coordinates
{
(0,0.47) (0.2,0.47)
(0.4,0.47) (0.6,0.47)
(0.8,0.47) (1,0.47)
};
\label{No fusion}

\addplot+[sharp plot,color=mycolor5, mark color=red, line width=1pt,mark=square*,mark options={fill=mycolor5}]
coordinates
{
(0,0.78) (0.2,0.68)
(0.4,0.46) (0.6,0.34)
(0.8,0.28) (1.0,0.27)
};
\label{F-Cooper}

\addplot+[sharp plot,color=mycolor3,line width=1pt,mark=pentagon*,mark options={fill=mycolor3}]
coordinates
{
(0,0.73) (0.20,0.64)
(0.40,0.49) (0.6,0.42)
(0.80,0.40) (1.0,0.38)
};
\label{OPV2V}

\addplot+[sharp plot,color=mycolor4,line width=1pt,mark=triangle*,mark options={fill=mycolor4}]
coordinates
{
(0,0.68) (0.2,0.53)
(0.4,0.36) (0.6,0.32)
(0.8,0.29) (1.0,0.29)
};
\label{Late fusion}

\addplot+[sharp plot,color=mycolor6,line width=1pt,mark=diamond*,mark options={fill=mycolor6}]
coordinates
{
(0,0.74) (0.2,0.61)
(0.4,0.38) (0.6,0.25)
(0.8,0.21) (1,0.2)
};
\label{Early fusion}

\addplot+[sharp plot,color=mycolor2,solid,line width=1pt,mark=*,mark options={fill=mycolor2}]
coordinates
{
(0,0.66) (0.2,0.64)
(0.4,0.63) (0.6,0.62)
(0.8,0.61) (1.0,0.60)
};
\label{Our model}

\end{axis}
\end{tikzpicture}}
\subfloat[Robustness with heading errors]{
\definecolor{mycolor1}{rgb}{0.00000,0.44700,0.74100}%
\definecolor{mycolor2}{rgb}{0.85000,0.32500,0.09800}%
\definecolor{mycolor3}{rgb}{0.92900,0.69400,0.12500}%
\definecolor{mycolor4}{rgb}{0.49400,0.18400,0.55600}%
\definecolor{mycolor5}{rgb}{0.46600,0.67400,0.18800}%
\definecolor{mycolor6}{rgb}{0.30100,0.74500,0.93300}%
\definecolor{mycolor7}{rgb}{0.63500,0.07800,0.18400}%
\definecolor{lightgray204}{RGB}{204,204,204}

\begin{tikzpicture}[font=\scriptsize]
\begin{axis}[
width=2.3in,
height=1.8in,
legend pos=outer north east,
legend style={fill opacity=0.8, draw opacity=1, text opacity=1, draw=lightgray204
},
ymin=0.2,ymax=0.8,
xmin=0,xmax=2.5,
xlabel=Heading error std $\sigma_\phi$ (\degree),
ylabel=Average precision at IoU 0.7,
ymajorgrids=true,
xmajorgrids=true,
grid style=dashed,
xtick={0,0.5,1,1.5,2.0,2.5}
]

\addplot[dashed,color=mycolor1,line width =0.75pt]
coordinates
{
(0,0.47) (0.5,0.47)
(1,0.47) (1.5,0.47)
(2.0,0.47) (2.5,0.47)
};

\addplot+[sharp plot,color=mycolor5, mark color=red, line width=1pt,mark=square*,mark options={fill=mycolor5}]
coordinates
{
(0,0.78) (0.5,0.65)
(1,0.50) (1.5,0.41)
(2.0,0.37) (2.5,0.33)
};

\addplot+[sharp plot,color=mycolor3,line width=1pt,mark=pentagon*,mark options={fill=mycolor3}]
coordinates
{
(0,0.73) (0.5,0.64)
(1,0.55) (1.5,0.50)
(2,0.46) (2.5,0.44)
};

\addplot+[sharp plot,color=mycolor4,line width=1pt,mark=triangle*,mark options={fill=mycolor4}]
coordinates
{
(0,0.68) (0.5,0.58)
(1,0.47) (1.5,0.40)
(2.0,0.36) (2.5,0.33)
};

\addplot+[sharp plot,color=mycolor6,line width=1pt,mark=diamond*,mark options={fill=mycolor6}]
coordinates
{
(0,0.74) (0.5,0.59)
(1.0,0.45) (1.5,0.37)
(2.0,0.33) (2.5,0.29)
};

\addplot+[sharp plot,color=mycolor2,solid,line width=1pt,mark=*,mark options={fill=mycolor2}]
coordinates
{
(0,0.66) (0.5,0.65)
(1.0,0.65) (1.5,0.64)
(2.0,0.63) (2.5,0.61)
};

\end{axis}
\end{tikzpicture}}
\subfloat[AP$70$ vs bandwidth requirement]{
\definecolor{mycolor1}{rgb}{0.00000,0.44700,0.74100}%
\definecolor{mycolor2}{rgb}{0.85000,0.32500,0.09800}%
\definecolor{mycolor3}{rgb}{0.92900,0.69400,0.12500}%
\definecolor{mycolor4}{rgb}{0.49400,0.18400,0.55600}%
\definecolor{mycolor5}{rgb}{0.46600,0.67400,0.18800}%
\definecolor{mycolor6}{rgb}{0.30100,0.74500,0.93300}%
\definecolor{mycolor7}{rgb}{0.63500,0.07800,0.18400}%
\definecolor{lightgray204}{RGB}{204,204,204}

\begin{tikzpicture}[font=\scriptsize]
\begin{axis}[
width=2.3in,
height=1.8in,
legend pos=outer north east,
legend style={fill opacity=0.8, draw opacity=1, text opacity=1, draw=lightgray204},
ymin=0.2,ymax=0.8,
xmode=log,
xlabel=Bandwidth requirement in log scale (Mbps),
ylabel=Average precision at IoU 0.7,
ymajorgrids=true,
xmajorgrids=true,
grid style=dashed,
]

\addplot+[dashed,color=mycolor5, mark color=red, line width=0.5pt,mark=square*,mark options={fill=mycolor5}]
coordinates
{
(72.08,0.78) (72.08,0.27)
};

\addplot+[dashed,color=mycolor3,line width=0.5pt,mark=pentagon*,mark size=3pt,mark options={fill=mycolor3}]
coordinates
{
(126.8,0.73) (126.8,0.38)
};

\addplot+[dashed,color=mycolor4,line width=0.5pt,mark=triangle*,mark size=3pt,mark options={fill=mycolor4}]
coordinates
{
(0.024,0.68)  (0.024,0.29)
};

\addplot+[dashed,color=mycolor6,line width=0.5pt,mark=diamond*,mark size=3pt,mark options={fill=mycolor6}]
coordinates
{
(7.68,0.74)  (7.68,0.2)
};

\addplot+[dashed,color=mycolor2,solid,line width=0.5pt,mark=*,mark options={fill=mycolor2}]
coordinates
{
(0.024,0.66) (0.024,0.60)
};

\addplot+[sharp plot,color=mycolor5, mark color=red, only marks, mark size=3pt,mark=square*,mark options={fill=mycolor5}]
coordinates
{
(72.08,0.78)
};

\addplot+[sharp plot,color=mycolor3,only marks, mark size=4pt,mark=pentagon*,mark options={fill=mycolor3}]
coordinates
{
(126.8,0.73)
};

\addplot+[sharp plot,color=mycolor4,only marks, mark size=5pt,mark=triangle*,mark options={fill=mycolor4}]
coordinates
{
(0.024,0.68)
};

\addplot+[sharp plot,color=mycolor6,only marks, mark size=5pt,mark=diamond*,mark options={fill=mycolor6}]
coordinates
{
(7.68,0.74)
};

\addplot+[sharp plot,color=mycolor2,solid,only marks, mark size=3pt,mark=*,mark options={fill=mycolor2}]
coordinates
{
(0.024,0.66)
};
\end{axis}

\node at (1.35,2.5) {\textbf{Without error}};

\node at (1.35,0.5) {With error};
\end{tikzpicture}
}
\vspace{-1mm}
\subfloat{
\definecolor{mycolor1}{rgb}{0.00000,0.44700,0.74100}%
\definecolor{mycolor2}{rgb}{0.85000,0.32500,0.09800}%
\definecolor{mycolor3}{rgb}{0.92900,0.69400,0.12500}%
\definecolor{mycolor4}{rgb}{0.49400,0.18400,0.55600}%
\definecolor{mycolor5}{rgb}{0.46600,0.67400,0.18800}%
\definecolor{mycolor6}{rgb}{0.30100,0.74500,0.93300}%
\definecolor{mycolor7}{rgb}{0.63500,0.07800,0.18400}%

\begin{tikzpicture}[font=\scriptsize]
\node[draw=white,outer sep=0.5pt,inner sep=0pt] at (0,0) 
{\scriptsize
     \begin{tabular}{llllllllllll}
     \ref{Our model}& Proposed model & 
    \ref{No fusion} & No Fusion &
    \ref{Early fusion}&Early fusion 
    \ref{Late fusion}& Late fusion &
    \ref{OPV2V}&OPV2V & \ref{F-Cooper}&F-Cooper 
    \end{tabular}
    };
\end{tikzpicture}}

\caption{Robustness assessment on dataset {Digital Culver City}}
\label{experiments:robustness assessment}
\end{figure*}

\begin{table*}[htbp]
    \centering
    \caption{AP$70$ on {OPV2V-Test} under different noise levels}
    \setlength{\tabcolsep}{3.5mm}    
    \begin{tabular}{@{}ccccccccccccc@{}}
\toprule
\doublerulesepcolor{red}
{$\sigma_p$(m)}& {0}&{0.2}&{0.4}& {0.6}&{0.8}&{1.0} & {$\sigma_\phi$($\degree$)}
&{0.5}&{1.0}&{1.5}&{2.0}&{2.5}\\
\midrule
No fusion
&0.60&0.60& 0.60& 0.60&0.60&0.60&No fusion&0.60& 0.60& 0.60&0.60&0.60\\
Early fusion
& \textbf{0.85}&0.72&0.40&0.25&0.19&0.17&Early fusion &0.72&0.54&0.42&0.36&0.30\\
Late fusion &0.80&0.60&0.34&0.24&0.23&0.25&Late fusion&0.64&0.47&0.38&0.32&0.29\\
F-Cooper 
&{0.82}&\textbf{0.74}&  0.49&0.32& 0.23&  0.19&F-Cooper& 0.69& 0.51&0.41&0.35&0.31\\
OPV2V 
&{0.82}&\textbf{0.74}& 0.58&0.49&0.44&0.42&OPV2V&0.74& 0.66&0.60&0.57&0.54\\
\textbf{ Our model}
&0.76 &\textbf{0.74} &\textbf{0.72} &\textbf{0.71} &\textbf{0.69} &\textbf{0.68}& \textbf{ Our model}& \textbf{0.75}& \textbf{0.74}&\textbf{0.73}& \textbf{ 0.72}&\textbf{0.70}\\
\midrule

\end{tabular}
    \label{tab:result on test_sigmap}
\end{table*}

\subsection{Equality of Transform Estimation}

 Thanks to the ground truth data in the OPV2V dataset, the estimated transform can be directly evaluated, helping us to explain the effectiveness of the proposed model. We decouple and evaluate the rotation and translation components of the ground truth transform $\mathcal{F}$ and the estimated one $\hat{\mathcal{F}}$ separately. 
 
 A rotation matrix $\mathbf{R}\in SO(3)$ can be expressed by the matrix exponential map $\mathbf{R}=\exp ([\mathbf{r}]_\times)$, and inversely $||\mathbf{r}||=\mathrm{arccos}\left(0.5\cdot \mathrm{Tr}
(\mathbf{R})-0.5 \right)$.
Then the ground truth rotation $\mathbf{R}$ and estimated rotation $\hat{\mathbf{R}}$ can be compared in the vector form with relative rotation error (RRE),       
\begin{equation}
\text{RRE} =||\mathbf{r}-\hat{\mathbf{r}}||= \mathrm{arccos}\left(0.5 \cdot {\mathrm{Tr}\big(
\mathbf{R}^T \cdot \hat{\mathbf{R}}\big)-0.5} \right).
\end{equation}
Similar to the evaluation of the rotation matrix, a relative translation error (RTE) is defined as
     \begin{equation}
         \text{RTE}=||\mathbf{t}-\hat{\mathbf{t}}||.
     \end{equation}

Fig. \ref{fig:rreandrte} illustrates the instantaneous curve of RRE and RTE tested on {Digital Culver City} with position error $\sigma_p=1$m. In the vanilla late fusion approach, RRE is absent as we solely introduced translation error for evaluation purposes to demonstrate the strengths and weaknesses of our proposed method. 
It is interesting to observe that lower RTE is achieved for the proposed model with a slightly larger RRE ($\leq 1^o$). 
Although small rotation errors are introduced, the estimated transformation matrices are accurate enough for cooperative perception with position and heading errors based on the numerical studies. 

\subsection{Robustness Assessment of Average Precision} 
The evaluation metric is the average precision (AP) by comparing the Intersection over Union (IoU) of fused bounding boxes and the ground truth boxes \cite{padilla2020survey}. We choose IoU=$0.7$, which means only the two boxes whose area of overlap exceeds $ 70\%$ of the area of union will be regarded as a True Positive detection. The proposed model and benchmarks are evaluated on two subsets of OPV2V.
 
\subsubsection{{Digital Culver City}}

Fig.\ref{experiments:robustness assessment}(a) and (b) show the results on {Digital Culver City}.  
The Ego vehicle senses the environment independently for no fusion case, therefore, the average precision of Intersection-over-Union $0.7$ (AP$70$) is a constant and keeps stable at $0.47$. 
For accurate location and heading scenarios  (\ie, $\sigma_p=0$m, $\sigma_\phi=0\degree$), early fusion and intermediate fusion models  (\eg F-Cooper and OPV2V) outperform late fusion, because of the less information loss. 
Meanwhile, similar performance is observed for the proposed method and the late fusion model.
But the performance of early fusion, late fusion, OPV2V, and F-Cooper all decrease dramatically when location and/or heading errors occurred.
The performance of F-Cooper drops below the no-fusion method when
 $\sigma_p \geq 0.4 $m or $\sigma_\phi \geq  1 \degree$.
While best performance is achieved for the proposed OT-based methods, for which only $6\%$ and $5\%$ decrease with respect to the noiseless case when $\sigma_p=1$m and $\sigma_\phi=2.5 \degree$. 

\subsubsection{OPV2V-Test}
 
This subset is utilized to evaluate the robustness, adaptability, and portability of the proposed method.
Table \ref{tab:result on test_sigmap} shows the AP$70$ of the models under different levels of $\sigma_p$ and $\sigma_\phi$. Note that the AP is higher than {Digital Culver City} because of the higher similarity to the training data. 
The performance of the benchmarks is similar to the previous test set, the proposed methods outperform the benchmarks by a large margin. It proves that our model works well on different types of datasets.

\subsection{Bandwidth Requirements}
Besides the robustness against spatial errors, the communication burden or bandwidth requirement is also an important factor for cooperative perception.  Fig. \ref{experiments:robustness assessment}(c) shows the bandwidth requirements and corresponding AP for the benchmarks and the proposed model. The bandwidth requirement is obtained by estimating the data size of transmitting information, \ie bounding boxes and classification confidence for late fusion, feature map  for intermediate fusion, and raw data size (point cloud) for early fusion. 
As an illustrate, for early fusion working at a frame rate of $f_r$ with $n_p$ points in the raw data, suppose the dimension of each point is $n_d$, and $n_b$ bits are required to describe each dimension, then the bandwidth requirement is calculated as

\begin{equation}
\text{BW}= f_r \times n_p \times n_d \times n_b \ (\text{bps})
\end{equation}

The bandwidth requirement of our model is approximately equal to late fusion because of the demands for only bounding boxes (typically less than 20) along with their confidence, which can be represented by an 8-D vector. 
Hence, the bandwidth requirement of our model is generally less than 
$51.2$Kbps, which outperforms the intermediate methods and early fusion by a huge margin. Here we calculate the bandwidth requirement for intermediate fusion using the uncompressed data size, since the process of feature map compression 1) is specific to corresponding fusion model, 2) needs a decoder at the receiver end highly coupled with an encoder at the sender end, 3) introduces considerable time delay in the processing procedure.

\section{Conclusions}
\label{sec:conclusions}
In this paper, we proposed a distributed cooperative perception framework working at the object level for connected and automated driving.
The inaccurate relative transformation caused by the position and pose errors is regarded as one of the main factors that affect the continuous robust mapping of the Ego vehicle. To address such a challenge, an optimal transport theory-based method is developed to find the correspondence between the measurements of both the Ego vehicle and the cooperative CAV, followed by estimating a correction relative transform from the matched object pairs. The observations of the cooperative CAVs are transformed into the Ego frame and fused by the Ego via Non-maximum suppression according to classification confidence. 
Experiments show that the proposed method outperforms the state-of-the-art performance on average precision when location or heading errors occur. Furthermore, the least bandwidth requirement is observed when compared to the benchmark models. The proposed method exhibits good generality, low cost, low communication burden, and ease of implementation.

\section*{Acknowledgement}
This work was supported in part by the National Key R$\&$D Program of China under Grant 2021YFB1600402 and 2020YFB1600303 and in part by Tsinghua University-Toyota Joint Center.


\end{document}